# Decoding Cellular Temperature via Neural Network-Aided Fluorescent Thermometry

Tong Zhang, Tian-Tian Li, Jing-Ru Wang, Yu-Wen Zhang, Chao Sun, Zheng Huang, Jing-Juan Xu, Bin Kang*

State Key Laboratory of Analytical Chemistry for Life Science, Chemistry and Biomedicine Innovation Center, School of Chemistry and Chemical Engineering, Nanjing University, Nanjing, 210023, China

*Corresponding author. Email: binkang@nju.edu.cn (B. K.)

## Abstract

The temperature distribution within cells, especially the debates on mitochondrial temperature, has recently attracted widespread attention[1-3]. Some studies have claimed that the temperature of mitochondria can reach up to 50-53 °C[4-6]. Yet others have questioned that this is due to measurement errors from fluorescent thermometry caused by other factors[7,8], like cell viscosity[9]. Here we presented a neural network-aided fluorescent thermometry and decoupled the effect of cellular viscosity on temperature measurements. We found that cellular viscosity may cause significant deviations in temperature measurements. We investigated the dynamic temperature changes in different organelles within the cell under stimulation and observed a distinct temperature gradient within the cell. Eliminating the influence of viscosity, the upper limit of mitochondrial temperature does not exceed 42-43 °C，supporting our knowledge about the inactivation temperature of enzymes[10]. The temperature of mitochondria is closely related to their functions and morphology, such as fission and fusion. Our results help to clarify the question on of "how hot are mitochondria?" and promotes better understanding on cellular thermodynamics.

# Main

Temperature regulation and heat balance are critical for maintaining life activities within organisms and cells[11-13]. Living organisms and cells have multiple regulatory mechanisms to ensure robust temperature adaptability[14,15]. The temperature distribution within cells has recently garnered significant attention[16,17]. It is theoretically predicted to be difficult to maintain a significant temperature gradient within a cell[7]. Experimentally, it has been observed that certain regions within the cell, such as the nucleus and mitochondria, exhibit higher temperatures than the cytoplasm[18,19]. In particular, the temperature of the mitochondria has been reported to possibly exceed 50-53 °C[4-6]. While this result was quickly questioned as being caused by measurement errors from the fluorescent thermometry[7]. Such skepticism is principally reasonable, because in fluorescent thermometry, how to quantify and eliminate factors other than temperature, particularly cell viscosity, has always been a persistent challenge[9,20]. Thus far, "how hot are mitochondria?" still remains an open question.

We here addressed this challenge through a neural network-aided fluorescent cellular thermometry. We involved both temperature and viscosity in to a two-dimensional calibration matrix instead of the conventional one-dimensional temperature calibration curve. Leveraging neural networks and matrix operations, we decoupled the effects of temperature and viscosity on the measurements. After training and refining the coefficient matrix from solution, artificial simulated cells, and fixed cell, we achieved temperature mapping in cells with a precision of 0.05 °C. We observed a distinct temperature gradient within the cell, with the lysosome temperature being ~1 °C lower, while the mitochondria temperature is 1-4 °C higher than that of the cytoplasm. Stimulation of cellular heat production leads to dynamic increases in the temperature of cell organelles as well as the overall cell temperature. However, even with intense heat-generating stimulation, the upper limit of mitochondrial temperature did not exceed 42-43 °C. This result corroborates the finding that the enzymes in mitochondria lose their activity above 43 °C[10]. We also revealed that the mitochondrial temperature is closely associated with their fusion and fission morphology, thereby providing a perspective for studying mitochondrial function and cellular thermodynamics.

# Result

**Thermosensitive probe and measurement principle**

The measurement relies on a ratiometric fluorescent probe comprising two molecular modules, wherein hexamethylenediamine links rhodamine to hemicyanin, as depicted in Fig. 1a. The module rhodamine serves as a typical temperature-sensitive dye as the diethylamino groups undergo rotational changes in response to environmental variations due to motion-induced changes in emission (MICE) effect[9] (see Supplementary Note 5 for detailed description). The module hemicyanin is utilized as a reference to mitigate measurement errors arising from concentration fluctuation and fluorescence quenching. These two molecular modules, with different emission wavelengths, can provide up to four measurement parameters, including two emission peak intensities and two lifetimes. We defined two parameters: one is the ratio of the emission peak intensities of the two modules ($R = I_1/I_2$), and the other is the lifetime of rhodamine ($\tau$). In principle, both of these measurement parameters are affected by temperature and viscosity, and it is nearly impossible to isolate each of them from traditional one-dimensional response curve. Here, by constructing a two-dimensional response matrix containing temperature and viscosity and combined with a neural network, we decouple the effects of temperature and viscosity on fluorescence. Please see Supplementary Fig. 4-6 for details regarding the spectral characterization of the probe and the construction of the response matrix.

To decouple these two variables (R and $\tau$) related to both temperature and viscosity, we utilized a neural network methodology for processing higher-dimensional data inputs and outputs. The weights within the neural network dictate the flow and intensity of data (Fig. 1b). In current study, the measurement variables (R and $\tau$), serves as the input data for the neural network. The coefficient matrix is derived from the measured output matrix (Supplementary Note 7). The coefficient matrix corresponds to the weights in the neural network, which reflects the relationship between input and output data. The matrix composed of temperature and viscosity is the output. (Fig. 1c). In the initial assignment of the neural network, the output matrix can be obtained through the operation of the input matrix and the coefficient matrix：

$$\begin{bmatrix} T \\ \eta \end{bmatrix} = [Matrix] \begin{bmatrix} \tau \\ R \end{bmatrix}$$

where $T$, $\eta$, $\tau$, $R$ is temperature, viscosity, fluorescence lifetime, and ratio respectively, *Matrix* illustrates the interrelationship among these four parameters.

To obtain the temperature and viscosity distributions in different regions within the cell, the aforementioned operations were performed pixel-to-pixel. Both fluorescence intensity as well as fluorescence lifetime signals of cells were captured (Supplementary Fig. 7). The fluorescence ratio R-mapping of cells can be readily obtained by dividing the fluorescence intensity in two channels ($I_{590}/I_{720}$). Furthermore, lifetime $\tau$-mapping of cells at 590 nm are obtained using a time-correlated single-photon counting (Supplementary Fig. 8). The R-mapping and $\tau$-mapping of cells then undergo the matrix operations pixel-to-pixel to produce the final mapping of both temperature and viscosity within cells (Fig. 1d). For the possible effects of other factors on the fluorescent probe, such as pH, ionic strength, thermal stability, and cytotoxicity, please see Supplementary Note 6 for details.

**The evolution and correction of coefficient matrix**

Although most similar studies previously would directly apply the calibration curves obtained in solution to cells[21,22], this approach would inevitably introduce unknown errors, as the intracellular environment is significantly different from that of a solution. The most ideal scenario would be to create an environment outside the cell that is completely identical to the intracellular environment for calibration curves, which is virtually impossible to achieve. To approximate the environment of living cells as closely as possible, we utilized the initial assignment, evolution, and correction of neural networks[23,24] to calibrate the coefficient matrix step-by-step from solution, to simulated cells, and finally to cells (Fig. 2a).

To begin with, aqueous glycerol solution samples at various temperatures and viscosities are employed to validate the accuracy initial coefficient matrix. The absolute errors in temperature and viscosity fall within an acceptable range (±0.08 °C; ±0.024 mPa·s) (Fig. 2b). Subsequently, giant plasma membrane vesicles (GPMVs) isolated directly from living cells (Supplementary Note 7), with more similar physiochemical properties to cell than aqueous solution, are used as a calibration model for measuring temperature. At different setting temperatures (27 °C, 32 °C, 37 °C, and 42 °C), GPMVs show a uniform inner temperature distribution close to the ambient temperatures, since the vesicle is almost entirely filled with cytoplasm and does not contain any organelles[25] (Fig. 2c). The temperature measurement error is approximately 0.1–0.3 °C, with an average of 0.22 °C (Supplementary Fig. 14). This value is larger than the 0.08 °C observed in the solution, clearly indicating that the change in environment has reduced the applicability of the coefficient matrix, leading to increased error. Through the evolution and correction of the coefficient matrix (see Supplementary Note 7 for details), the temperature measurement accuracy converged back to 0.06 °C (Fig. 2e).

Next, we applied the calibrated coefficient matrix from the GPMVs to fixed cells and found a deviation of ~0.5 °C within the setting temperature range of 27–42 °C (Fig. 2d). This deviation likely originates from two aspects: first, the complex intracellular environment containing various organelles cannot be simply regarded as a solution or homogeneous cytoplasm[26]. Second, the autofluorescence of the cells themselves may interfere with the measurements[27,28]. After correction of the coefficient matrix, the temperature measurement accuracy within the cells also reached ±0.05 °C (Fig. 2f). Although fixed cells contain heterogeneous organelle structures, the temperature distribution inside is still uniform and very close to the ambient temperature. This is quite reasonable, because fixed cells lose their metabolic activity and no longer have active heat-production. Once the cells reach heat exchange equilibrium with the environment, the temperature distribution inside becomes uniform. Due to the active heat generation and heterogeneous temperature distribution within living cells[29,30], the true value of temperature at a specific region is unknown in advance. Therefore, here the fixed cells represent the most ideal environment for calibration that can be achieved. The above results clearly demonstrate the risks associated with fluorescence thermometry in cells. Using calibration curves established in solution may lead to significant measurement deviations for cellular measurements.

**The temperature and viscosity within cells**

After calibrating the coefficient matrix, we measured the temperature and viscosity distributions within fixed and living cells. As the ambient temperature increased from 27 °C to 42 °C, the temperature inside the fixed cells changed synchronously with the ambient temperature, always showing a very uniform distribution (Fig. 3a). Compared with fixed cells, the temperature inside living cells exhibited significant heterogeneity (Fig. 3b). The temperature distribution range for living cells can be considerably broader, ~4.6 °C, in contrast to the narrower distribution range observed in fixed cells, which typically averages ~0.47 °C. (Fig. 3c, 3d). Another prominent distinction lies in the difference between their average cellular temperature and the ambient temperature. The average temperature of living cells exceeds ambient temperatures by approximately 0.6-0.8 °C, while the average temperatures for fixed cells always closely align with that of ambient environment (Fig. 3e). The above results demonstrate the unique characteristics of temperature distribution within living cells. First, there is a significant temperature heterogeneity of several degrees between different regions within the living cell. Second, there is a non-negligible temperature gradient between the intracellular temperature and the ambient temperature. This indicates an active heat generation within the cell and heat dissipation towards the environment.

Moreover, as the ambient temperature changes, the viscosity within fixed and living cells exhibits different trends. The viscosity within fixed cells decreased continuously with the increase in ambient temperature. In contrast, the viscosity within living cells hardly changed. (Fig. 3a, Fig. 3f). The interior of fixed cells can be regarded as a static biological material or medium, and the decrease in viscosity with increasing temperature is a common physical property of many media. However, living cells contain dynamic chemical reactions, molecular processes, and regulatory mechanisms, and their viscosity-temperature dependence cannot be considered as that of a simple material or medium. Clearly, living cells exhibit active regulatory behaviors, adjusting their internal viscosity in response to different ambient temperatures to ensure the robustness of the intracellular environment[14,31].

**Temperatures on different organelles**

The temperature profile of living cell clearly shows a temperature gradient of 4–5 °C inside the cell. When the ambient temperature is maintained at 37 °C, the average temperature recorded for the living cell is 37.8 °C, with the highest local temperature reaching 39.7 °C and the lowest at 35.7 °C (Fig. 3, Supplementary Fig. 17). We then decrypt the temperature differences among various organelles within the cell. The mitochondria are the first priority organelles we focused on, as they are the power station of the cell and the primary source of heat production[32,33]. The probe we employed is a cationic fluorescent dye characterized by its rhodamine structure, which carries a positive charge that facilitates its accumulation in mitochondria (Fig. 4a). Co-localization experiments with Mito-Tracker indicate that the probe partially targets the mitochondria (Supplementary Fig. 18). Additionally, when the probe is taken up by the cell or dynamically released from the mitochondria, it can also be captured by lysosomes. Co-localization experiments with LysoSensor show that a portion of the probe is distributed in lysosomes (Fig. 4a). Some of the probe, after detaching from the mitochondria or escaping from the lysosomes, is also distributed in the cytoplasm.

To further understand the temperature profile among different cellular organelles, we developed an image segmentation method based on co-localization imaging, image boundary recognition, and image registration. As shown in Fig. 4b, in brief, we label the mitochondria and lysosomes of the cell with Mito-Tracker and LysoSensor, respectively, and extracted the boundaries of the organelle images through image processing. Subsequently, we registered the boundaries of each organelle with the fluorescence imaging of the thermosensitive probe, identified and extracted the overlapping regions. Through this process, we can deconvolve the temperature distribution map of the cell into separate maps for mitochondria, lysosomes, and the cytoplasm. Noted that during the image registration process, some data points may be lost. Some regions of the mitochondria are discarded due to poor co-localization. Some regions containing the thermosensitive probe could not be labeled by either Mito-Tracker or LysoSensor, and the fluorescence appeared as diffuse distribution without a clear organelle morphology. We consider these regions to likely represent the cytoplasm.

Next, we investigated the temperature changes in different organelles within the cell under resting conditions and upon heat production induced by $Ca^{2+}$ shock stimulation. The $Ca^{2+}$ shock stimulation is induced using ionomycin calcium salt, which increases intracellular calcium ion levels and alters the rate of respiratory reactions[34]. As the concentration of calcium ions rises, the cell temperature increases significantly following treatment with 1.0 μmol of ionomycin calcium salt (Fig. 4c). Higher concentrations of ionomycin calcium salt (1.5 μmol), however, led to a decrease in cell temperature, which may be due to excessive $Ca^{2+}$ shock damaging the mitochondria and thereby affecting cell heat production. Fig. 4d and e show the temperature distribution on the organelles in the cell under resting conditions and after treatment with 1.0 μmol of ionomycin calcium salt. When the ambient temperature is set at 37 °C, the average temperature of the mitochondria is 39.0 °C, which is significantly higher than the temperature of the cytoplasm at 37.6 °C. Interestingly, we note that the temperature of the lysosomes is only 36.7 °C, approximately 0.8-1 °C lower than the cytoplasmic temperature. This suggests a process within the cell where heat is produced by the mitochondria, transferred to the cytoplasm, and ultimately dissipated into the environment. However, we currently do not fully understand why the temperature of the lysosomes is lower than that of the cytoplasm. One possibility is that there are some endothermic hydrolytic reactions occurring within the lysosomes[35], or that the proton gradient created by the proton pumps on the lysosomal membrane affects the local temperature[36]. A complete understanding of the underlying mechanism will undoubtedly require more in-depth research.

Following treatment with 1.0 μmol of ionomycin calcium salt, there is a notable increase in cellular temperatures. The average temperatures for lysosomes and mitochondria rise to approximately 38.3 °C and 40.1 °C respectively, and the cytoplasmic temperatures also increase significantly to around 39 °C. The elevated concentration of $Ca^{2+}$ in cells could enhanced thermogenesis rates within mitochondria. It is evident that more heat is subsequently transferred to both lysosomes and cytoplasm. This result is consistent with previous observations of dynamic temperature transfer from the mitochondria to the lysosomes[37]. Here, even under the maximum calcium shock-induced heat production that we could achieve, the average temperature of the mitochondria only increased by approximately 1 °C. Even in regions of the mitochondria where local temperatures are relatively higher, the highest measured temperature do not exceed 42 °C. Some in vitro experiments have confirmed that important enzymes in the mitochondria can withstand a maximum temperature of around 43 °C[10]. Our results suggest that 42-43 °C is the upper limit of temperature that mitochondria can reach.

**Mitochondrial temperature and function**

The temperature of mitochondria and its regulatory mechanisms have also become a focus of widespread attention[38-40]. We then utilized the aforementioned image segmentation method to extract the temperature distribution map of mitochondria from cell imaging. Here we used two agents, carbonyl cyanide-p-trifluoromethoxyphenylhydrazone (FCCP) and rotenone, to stimulate and inhibit heat production on mitochondria, respectively. Specifically, FCCP stimulates cellular heat production by disrupting the proton gradient across the inner mitochondrial membrane, leading to impaired ATP synthesis and increased heat production[41]; Rotenone suppresses cellular heat production by inhibiting electron transport in complex I of the mitochondrial respiratory chain and disrupting mitochondrial respiration[42]. When the ambient temperature is maintained at 37 °C, following the addition of rotenone, the temperature decreased from 37.7 °C to 37.1 °C over a period of 15 minutes. In contrast, after adding FCCP, the average mitochondrial temperature rises from 37.7 °C to 39.8 °C within 15 minutes (Fig. 5a). Moreover, the gradual increase in FCCP concentration do not result in a significant alteration of mitochondrial temperature. Conversely, an excessively high concentration of FCCP decreased the mitochondrial temperature (Supplementary Fig. 22).

Mitochondria are highly dynamic organelles, and the delicate balance between fusion and fission processes is critical for maintaining mitochondrial homeostasis[43,44]. Typically, the fusion facilitates mitochondrial repair and enhances oxidative phosphorylation capacity. Conversely, excessive mitochondrial fission can lead to oxidative stress, reduced ATP production, and the release of pro-apoptotic proteins[45-47]. Mdivi-1 serves as a selective inhibitor of mitochondrial division, promoting mitochondrial elongation[48]. In contrast, Staurosporine (STS) induces fragmentation of mitochondria within cells[49] (Fig. 5b). Then the impact of mitochondrial morphology on their temperature is investigated through drug induction across four controlled trials (Fig. 5c). Compared to group I, which receives no drugs, mitochondria in group II (treated solely with Mdivi-1) exhibit significant aggregation and form a reticular morphology as observed via confocal imaging. This is further corroborated by enhanced mitochondrial fusion evident in transmission electron microscopy (TEM) images. In group III (exposed only to STS), mitochondria are distinctly fragmented into punctate structures under confocal imaging analysis while TEM revealed excessive division of these organelles. In comparison, the mitochondrial morphology observed in group IV (treated with both Mdivi-1 and STS) avoid extreme states of either fusion or fragmentation. It maintains a relatively balanced state akin to that seen in group I due to Mdivi-1's inhibition of STS-stimulated Drp1-dependent excessive division[46]. The mitochondrial areas across the four groups exhibits distinct differences. Group II has the largest area, while group III is the smallest. Groups I and IV show only slight variations (Fig. 5d-V).

We mapped the temperature with varying levels of mitochondrial aggregation. When the ambient temperature is at 37 °C, the average cellular temperature in group I is approximately 37.7 °C, the temperature in groups II and III are about 37.6 °C and 38.0 °C, respectively. This indicates that increased mitochondrial fusion reduces heat production, whereas fragmented mitochondria are associated with an increased heat production and higher temperature. Excessive division of mitochondria leads to a significant generation of reactive oxygen species (ROS), which can induce oxidative stress to cells[50,51]. This may contribute to excessive heat production observed in these cells. The average temperature of group IV is approximately 37.8 °C, which is lower than that of group III. This result suggests that Mdivi-1 may mitigate the cellular damage induced by STS treatment (Fig. 5e). To the best of our knowledge, this is the first time to observe the direct relationship between mitochondrial morphology (i.e., fusion or fission) and their temperature. This finding is enlightening for a deeper understanding of the functions of mitochondria within cells.

**Cellular thermal response to bacterial infection**

Many diseases are often accompanied by a fever, for instance, bacterial infections usually cause an increase in body temperature[52]. Despite the extensive research on bacterial infections, the thermal response and temperature change at single cell level to the infection processes is still unclear. Here we investigate the infection of cells by S. aureus, the one associated with the most deaths in adult among all infections[53]. The process of bacterial invasion is visualized using acridine orange (AO) staining. As a widely utilized and effective fluorescent dye, AO binds to the nucleic acids of cells or bacteria, emitting enhanced fluorescence[54]. Upon introducing S. aureus into the medium (Fig. 6a), fluorescent spots become visible as S. aureus approaches and attaches to the cell surface. Ultimately, S. aureus invades the interior of the cell (Fig. 6b). Higher concentrations of bacteria and prolonged infection duration lead to reduced cell viability (Supplementary Fig. 23). Following infection, cellular inflammation becomes markedly exacerbated, as evidenced by elevated levels of cytokines such as IL-6, IL-8, and TNF-α (Fig. 6c).

Then monodansylcadaverine (MDC) is employed to detect phagocytosis within cells. Notably, fluorescence intensity significantly increases over time during infection, indicating that lysosomal phagocytosis undergoes substantial changes in response to bacterial invasion (Fig. 6d). Cell temperature and viscosity are measured during the invasion of S. aureus. The ambient temperature is maintained at 37 °C, while the cell temperature is approximately 37.6 °C prior to the addition of S. aureus. Following the introduction of bacteria, the average cellular temperature increases to 39.6 °C within 10 minutes. Over the subsequent five-minute period, the cellular temperature decreases to 39.4 °C, and the cell progressively exhibits signs of mortality. Concurrently, cell viscosity also exhibits a time-dependent increase. Specifically, it rises from 2.9 mPa·s to 4.2 mPa·s over a period of 15 minutes (Fig. 6e). This result demonstrates the complex stress behavior of temperature and viscosity changes in cells when they are infected by bacteria. We observe that in the late stage of infection, the fluorescence of the probe dye gradually diffused from the cytoplasmic region (mainly the mitochondria) to the entire cell, including the nucleus. One possible reason is that the S. aureus contains a substantial amount of negatively charged teichoic acid, which has an affinity for attracting positively charged dyes through electrostatic interactions. Consequently, the probe dyes are transported into the nucleus along with bacteria. Another possible reason might be that bacterial reinfection disrupts nuclear structure integrity, thereby facilitating dye penetration (Supplementary Fig. 25).

## Discussion

Recently, there has been much debate surrounding the use of fluorescent thermometry to measure the temperature distribution within cells. The crux of the controversy lies in how to ensure that the measurement results are not caused by measurement errors. The underlying physicochemical rationale for this skepticism is that even for temperature-sensitive fluorescent probes like MTY, temperature is not the only affecting factor[55]. The viscosity heterogeneity within cells is a potential cause of fluorescence changes and may be mistakenly attributed to temperature variations. Therefore, the calibration curves established for fluorescent thermometers in solution must be applied to cells with great caution. Our analysis indicates that traditional calibration-curve methods, whether ratiometric or lifetime-based, may introduce significant deviations in temperature measurements. Please refer to the deviation analysis section in Supplementary Note 9.

We attempt to address this challenge from the fundamental logic of metrology. Instead of constructing a single-function response curve, we expand to a two-dimensional or even higher-dimensional response matrix. By associating experimental data with the initial values of labeling and then refining and correcting the matrix, we can decouple the effects of temperature and viscosity on fluorescence. This approach has enabled us to obtain a high-precision temperature distribution within cells, eliminating the interference of viscosity. Our results suggested an upper limit for mitochondrial temperature about 42-43 °C，supporting our knowledge about the inactivation temperature of enzymes[10]. We also found that the temperature of lysosomes is 2-6 °C lower than mitochondria and even ~1 °C lower than the cytoplasm, which is consistent with the previously reported temperature transfer from mitochondria to lysosomes[37]. Moreover, for the first time, we revealed a direct association between mitochondria temperature and their morphology. Fragmented mitochondria are associated with increased heat production and higher temperatures. These findings provide new insights into the study of thermobiology, including cellular adaptation and response to temperature, cell responses under extreme hot or cold environment, the regulation of cellular functions by temperature such as stem cell differentiation and neural remodeling.

## Methods

**Synthesis and characterization of thermosensitive probe**

Detailed procedures for the synthesis of probe are described in the Supplementary Note 2. After purification by semi-prepared HPLC, their chemical structures are characterized by mass spectroscopy (Supplementary Fig. 2).

**Fluorescence spectral and lifetime measurements**

The procedures and methods of fluorescence spectral and lifetime measurements are specifically described in Supplementary Note 4.

**Stability of fluorescent signal**

The fluorescence ratio and lifetime of the thermosensitive probe are measured under varying pH level and ionic strength. Detailed information regarding the regulation of solution pH and ionic strength can be found in Supplementary Note 6.

**Cell culture and probe incubation**

Details regarding cell culture and probe incubation are provided in Supplementary Note 8.

**Cell cytotoxicity assay**

The MTT assay is conducted to assess the cytotoxicity of thermosensitive probe on cells. Detailed descriptions of the experimental procedures can be found in Supplementary Note 6.

**Drug stimulation of cells**

The detailed experimental procedures examining the effects of drugs on cells are provided in Supplementary Note 12.

**S. aureus infection**

The detailed procedure of S. aureus infection is provided in Supplementary Note 13.

## Acknowledgements

This work was mainly supported by the National Natural Science Foundation of China (grant no. 22250009, 22261132510 and 22174064 to B. K.), Excellent Research Program of Nanjing University (grant no. ZYJH004 to B. K. and J.-J. X.).

## Author contributions

B. K., and J.-J. X. supervised the project. B. K. conceived the idea. T. Z. performed the experiments and analyzed the data. T.-T. L., J.-R. W. and Y.-W. Z contributed on experiments. C. S. contributed on data analysis. Z. H. contributed on fluorescent probes. B. K. and T. Z. wrote the initial manuscript. All authors discussed and commented on the manuscript.

## Competing interests

The authors declare no competing interests.

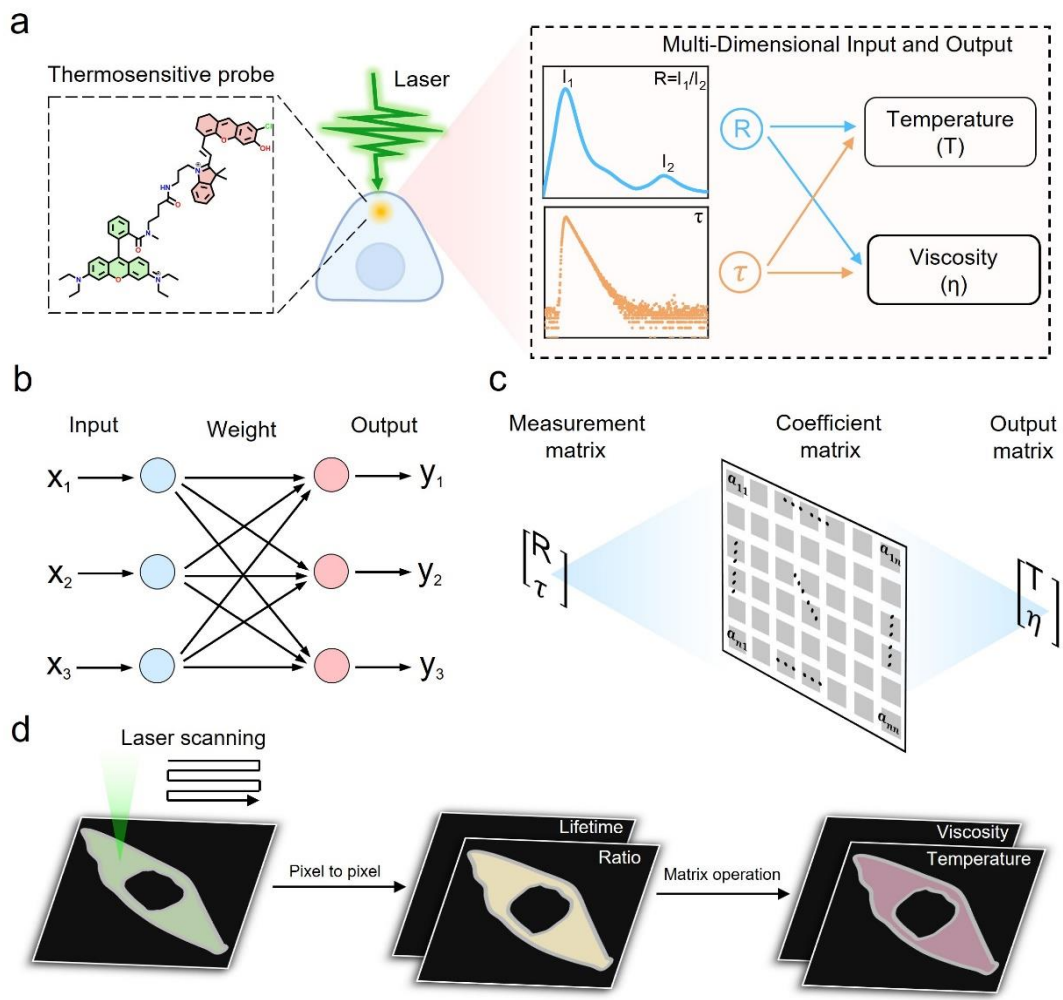

**Fig. 1 | Thermosensitive probe and measurement principle.** (a) Thermosensitive organic fluorophores are used to measure temperature in single cells through the response of fluorescence signals to temperature. As an additional environmental factor, viscosity can also influence the fluorescence signal. (b) Neural networks facilitate the simultaneous processing of multiple sets of data for both input and output. (c) The measurement matrix comprises fluorescence ratios and lifetimes, while the output matrix encompasses information on temperature and viscosity. The relationship between these two matrices is defined by the coefficient matrix. (d) Laser scanning with pixel to pixel allows for the acquisition of both R-mapping and τ-mapping for cells. Subsequently, all pixels are subjected to matrix operations to produce temperature and viscosity mapping for cells.

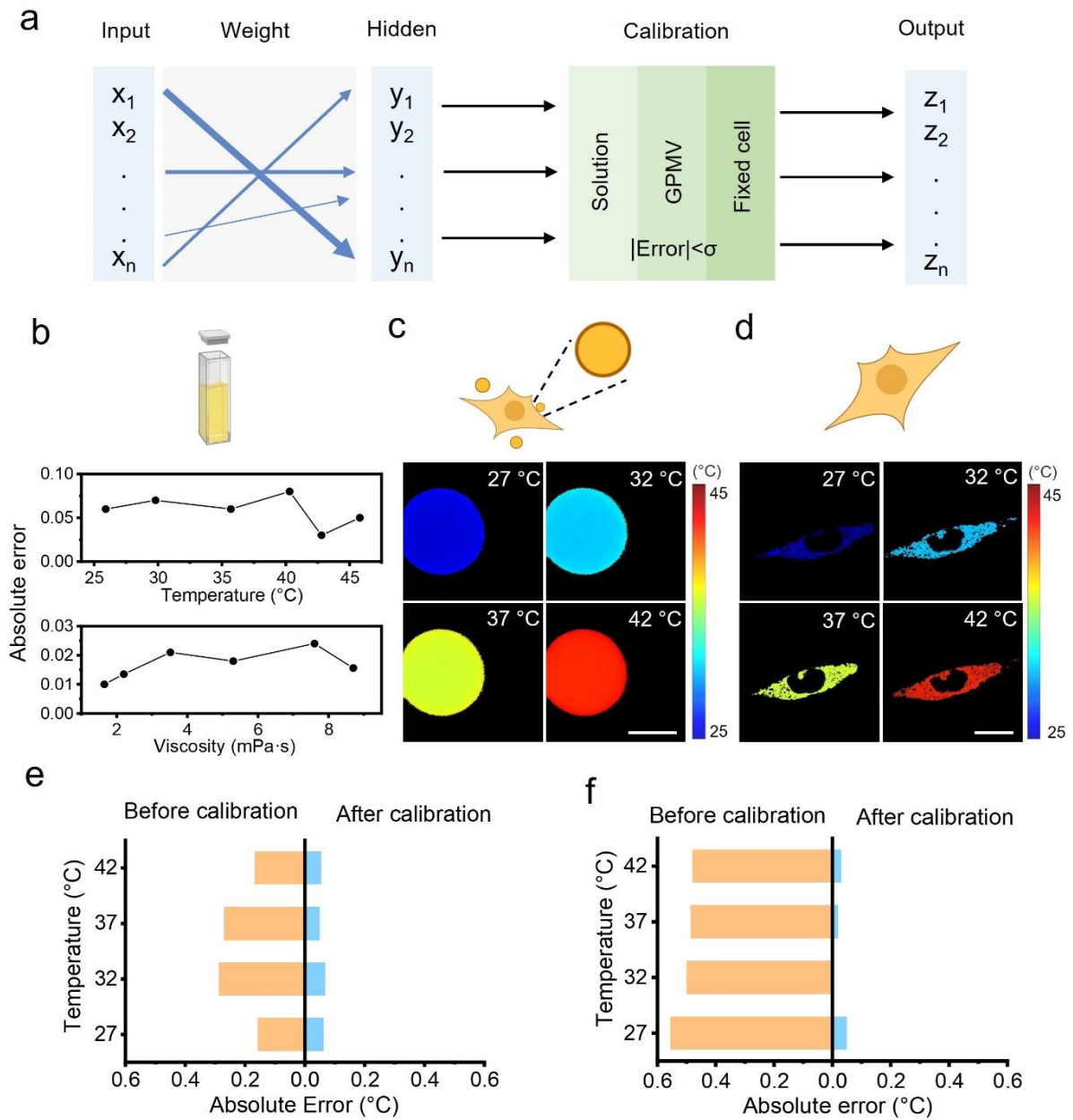

**Fig. 2 | The evolution and correction of coefficient matrix.** (a) The process of correcting coefficient matrix. After establishing the initial coefficient matrix in solution, calibrations are applied through solution, GPMVs, and fixed cells to enhance compatibility with cell thermometry. (b) Measurement errors of temperature and viscosity remains minimal in the solution. Temperature mapping of GPMVs (c) and fixed cells (d) under varied ambient temperature. For (c) and (d), the scale bars represent 1 μm and 5 μm, respectively. Temperature measurement errors in GPMVs (e) and fixed cells (f) before and after coefficient matrix calibration at different ambient temperatures.

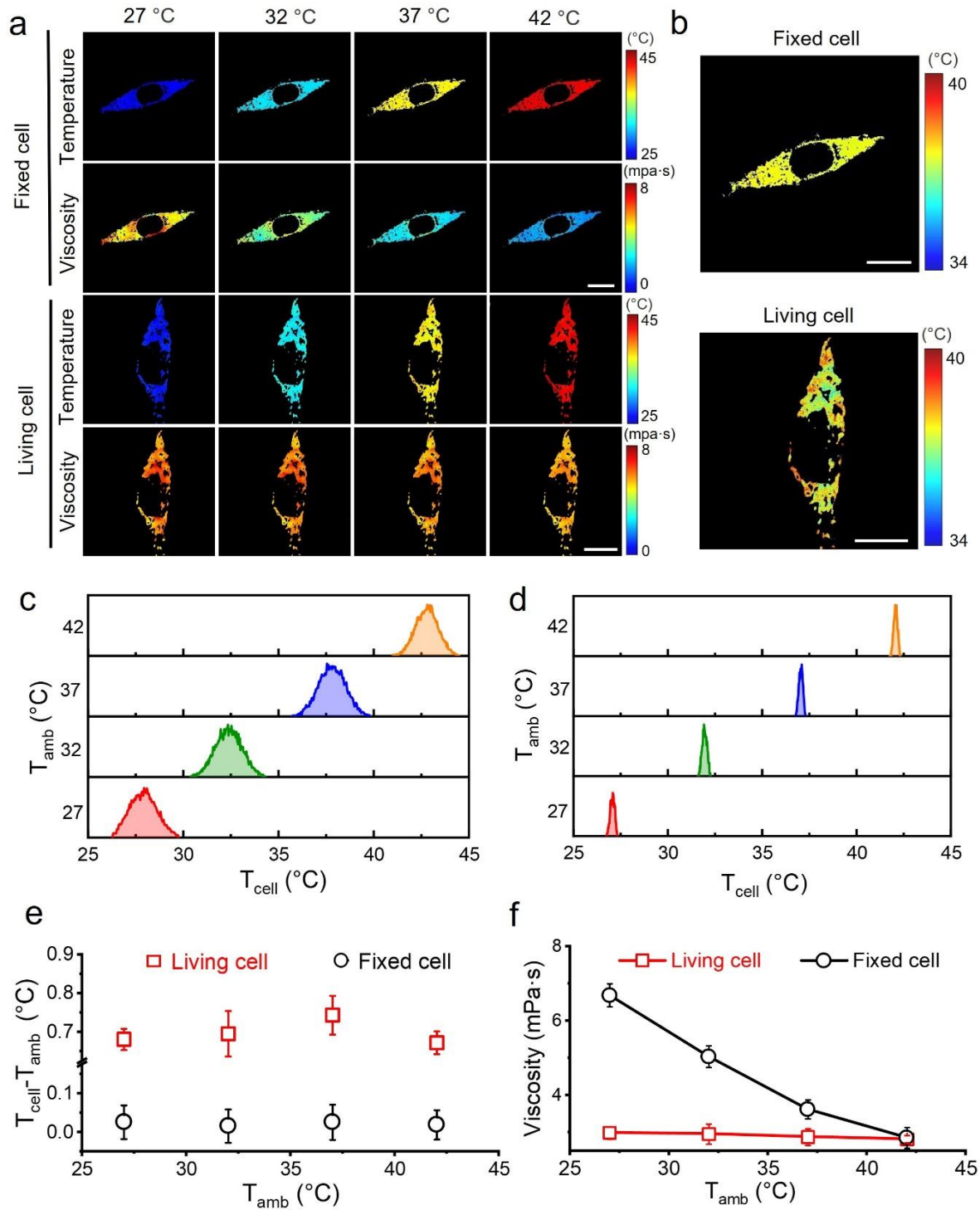

**Fig. 3 | Temperature and viscosity mapping in single cell.** (a) The temperature and viscosity imaging of both living and fixed cells under varying ambient temperatures. Scale bar represents 5 μm. (b) Temperature mapping for fixed cells and living cells at 37 °C. Scale bar represents 5 μm. Temperature profile for single living cells (c) and fixed cells (d) at varying ambient temperatures. (e) Temperature differences between cells and the environment. (f) Viscosity responses for fixed cells and living cells with varying temperature. $T_{amb}$ and $T_{cell}$ mean ambient temperature, average temperature for cell.

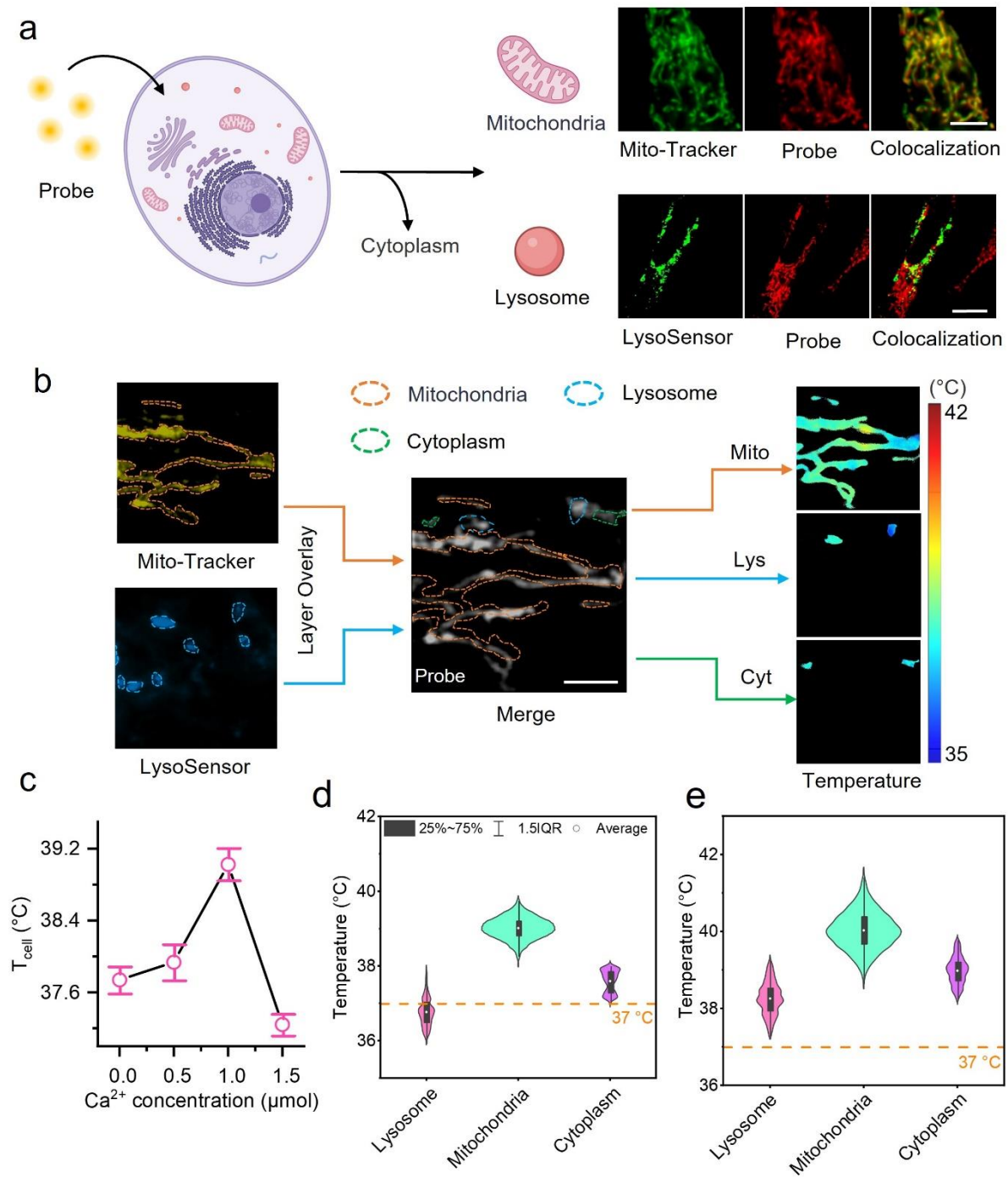

**Fig. 4 | Regional temperature difference in living cells.** (a) Distribution of thermosensitive probe. Scale bar represents 2 μm. (b) Schematic illustration of image segmentation method. Mito-Tracker, LysoSensor and thermosensitive probe layer overlap. Temperatures on mitochondria, lysosome and cytoplasm are extracted. Scale bar represents 2 μm. (c) The effect of $Ca^{2+}$ stimulation on cell temperature. Temperature profile of lysosome, mitochondria and cytoplasm with no $Ca^{2+}$ stimulation (d) and 1.0 μmol $Ca^{2+}$ stimulation (e).

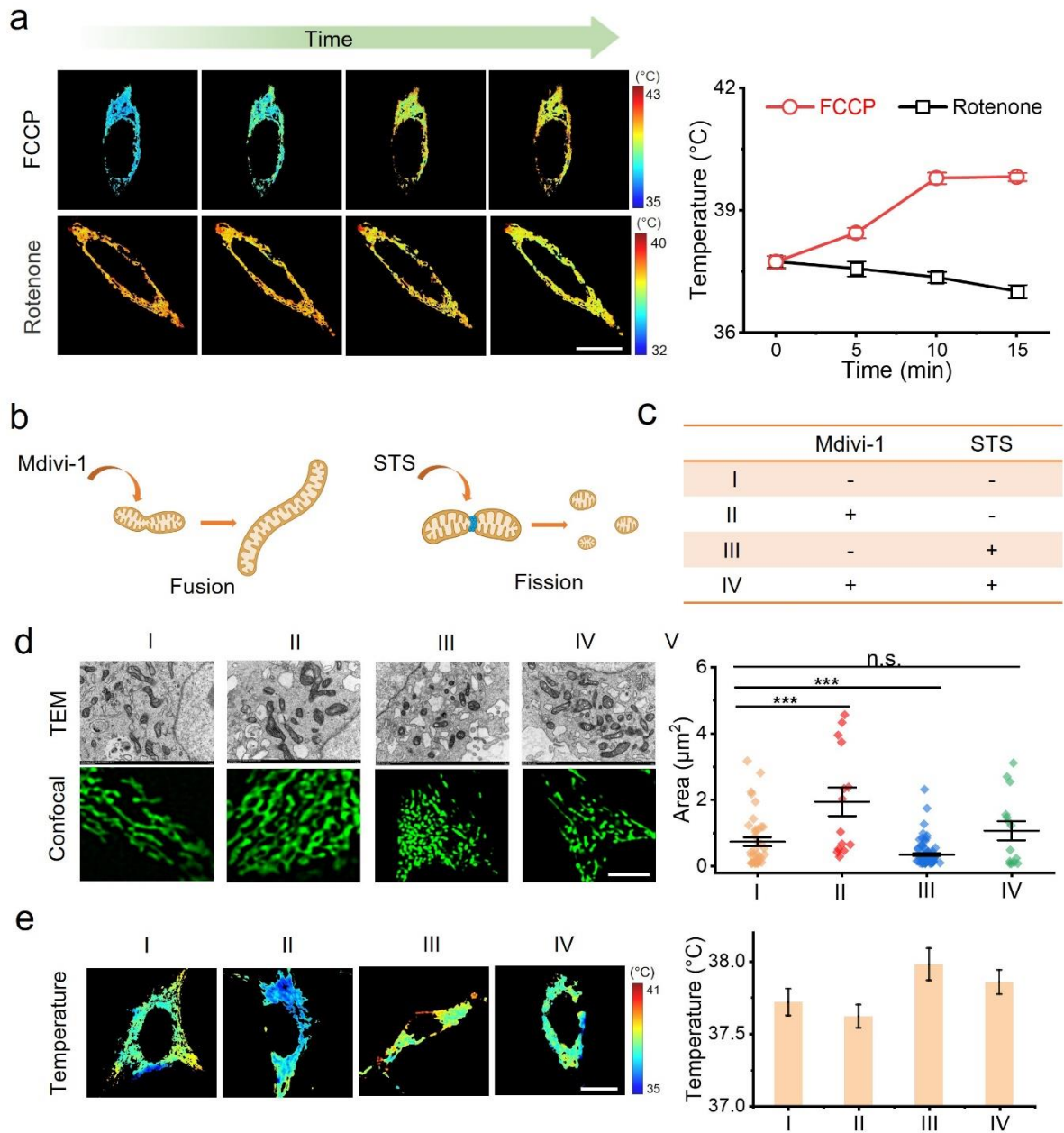

**Fig. 5 | Mitochondrial temperature and function.** (a) The temperature fluctuations of mitochondria following the addition of rotenone and FCCP. Scale bar represents 5 μm. (b) Schematic illustration of mitochondrial fusion and fission. Mdivi-1 induces mitochondrial fusion, while STS promotes mitochondrial fission. (c) Drug addition in four groups of controlled experiments. (d) Mitochondrial morphology and area in four controlled experiments. Mitochondrial area is analyzed via ImageJ. Scale bar represents 2 μm. (e) Temperature of mitochondria in four controlled experiments. *P < 0.05, **P < 0.01, and ***P < 0.001. Scale bar represents 5 μm.

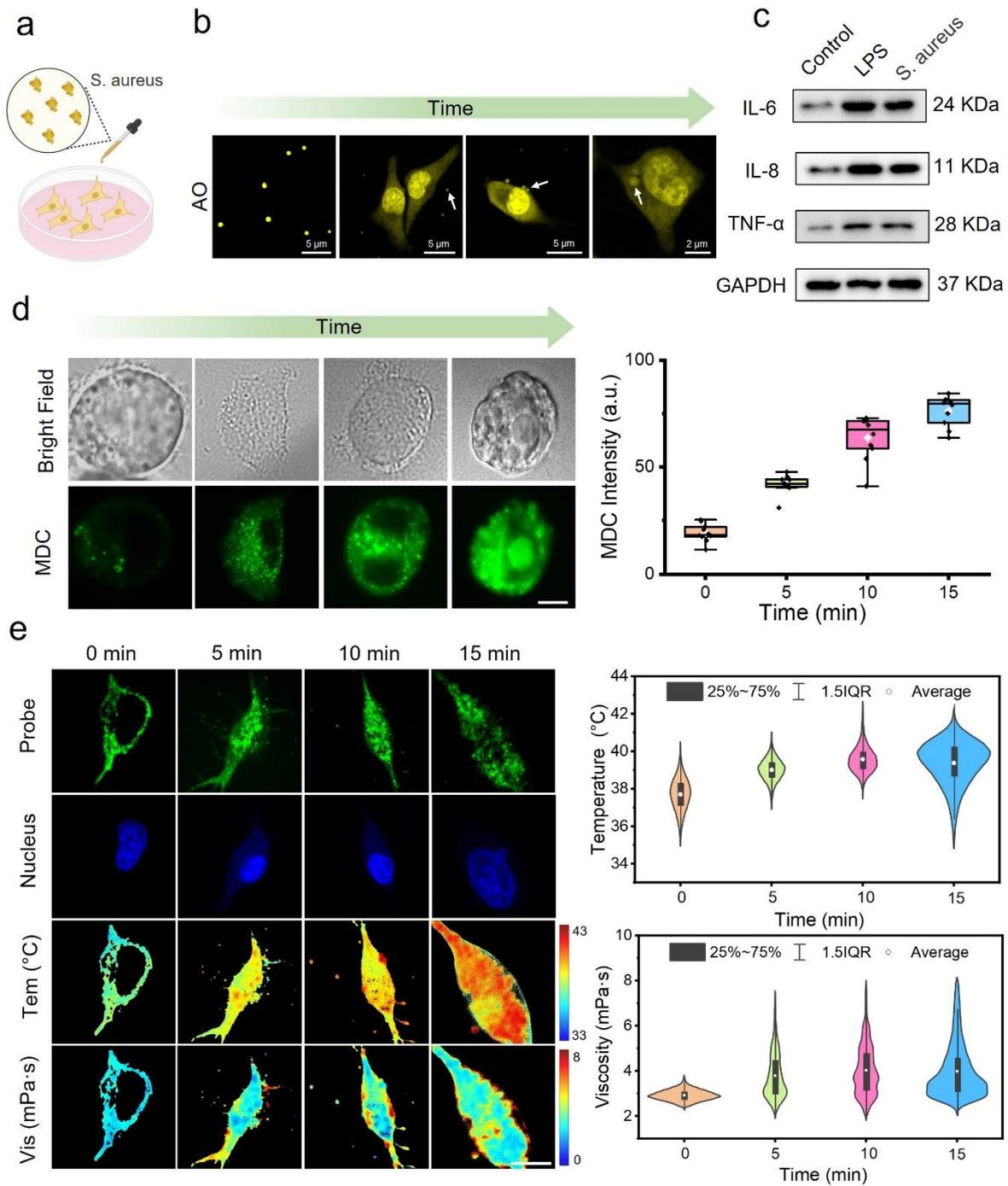

**Fig. 6 | Cellular thermal responses to bacterial infection.** (a) S. aureus is directly introduced into cell culture. (b) The process through which S. aureus invade Hela cells. (c) Infection triggers cellular inflammatory responses. The cytokines, such as IL-6, IL-8, and TNF-α, are elevated. (d) The MDC staining assay is employed to assess autophagy, revealing that the level of autophagy increased in response to bacterial infection. Scale bar represents 2 μm. (e) Cellular thermal and viscosity responses following S. aureus infection. Extensive cellular apoptosis occurs following a 15-minute duration. Scale bar represents 5 μm.